\documentclass{article}

\usepackage{microtype}
\usepackage{graphicx}
\usepackage{subcaption}
\usepackage{booktabs}

\usepackage{hyperref}

\usepackage[accepted]{icml2026}

\usepackage{amsmath}
\usepackage{amssymb}
\usepackage{mathtools}
\usepackage{amsthm}
\usepackage{bm}

\usepackage{tikz}
\usetikzlibrary{arrows.meta, positioning, shapes.geometric, fit,
                decorations.pathreplacing, calc, backgrounds}

\usepackage[capitalize,noabbrev]{cleveref}

\theoremstyle{plain}

\theoremstyle{definition}

\theoremstyle{remark}

\newcommand{\Msun}{\mathrm{M}_{\odot}}
\newcommand{\Halpha}{H\ensuremath{\alpha}}
\newcommand{\kms}{\,\mathrm{km\,s^{-1}}}
\newcommand{\Mdot}{\dot{M}}
\newcommand{\vinf}{v_{\infty}}

\icmltitlerunning{SBI of Colliding-Wind Binaries from \Halpha{} Time Series}

\begin{document}

\twocolumn[
  \icmltitle{
  Amortized Simulation-Based Inference of Colliding-Wind Binaries \\ from Short, Noisy Image Time Series}

  \icmlsetsymbol{equal}{*}

  \begin{icmlauthorlist}
    \icmlauthor{Niklas Knöll}{iwr,unihd}
    \icmlauthor{Tobias Buck}{iwr,unihd}
    \icmlauthor{Lorenzo Branca}{iwr,unihd}
    \icmlauthor{Giuseppe Viterbo}{iwr,unihd}
  \end{icmlauthorlist}

  \icmlaffiliation{iwr}{Interdisciplinary Center for Scientific Computing, Heidelberg, Germany}

  \icmlaffiliation{unihd}{University Heidelberg, Heidelberg, Germany}
  
  \icmlcorrespondingauthor{Tobias Buck}{tobias.buck@iwr.uni-heidelberg.de}

  \icmlkeywords{simulation-based inference, neural posterior estimation, normalizing flows, amortized inference, colliding-wind binaries, spatio-temporal encoding, astrophysical inverse problems, time-series inference}

  \vskip 0.3in
]

\printAffiliationsAndNotice{}

\begin{abstract}
Colliding-wind binaries (CWBs), which are systems of two massive stars whose supersonic winds collide into bow shocks, encode rich information about stellar wind properties in their multi-frequency emission, e.g. images in the \Halpha{}, X-ray, and radio wavelengths. Inferring physical parameters (mass-loss rates, terminal wind velocities, orbital elements) from short time-series observations is a compelling but challenging inverse problem, because the forward hydrodynamic simulator is computationally expensive and the likelihood is intractable. We adopt a factorized spatio-temporal architecture for amortized posterior inference that separates spatial encoding from temporal aggregation. This design aligns with the structure of the underlying physical process of local morphology and global dynamical evolution, induces time-translation equivariance in the learned representation, and improves identifiability in low-signal regimes. Coupled with a neural spline flow conditioned on these spatio-temporal embeddings of 10-frame \Halpha{} photon-count time series, we present a complete simulation-based inference pipeline for CWBs. Our method jointly infers seven physical parameters from synthetic observations under realistic detector noise, with posteriors verified as well-calibrated via TARP and SBC diagnostics. The approach naturally expands posterior width in information-poor regimes (low photon counts) and robustly recovers orbital parameters and mass-loss rates, demonstrating the feasibility of amortized likelihood-free inference for this challenging astrophysical inverse problem.
\end{abstract}

\section{Introduction}
\label{sec:intro}
Stellar winds are continuous outflows of mass, momentum, and energy from a star into its surrounding medium, and the dominant driving mechanism depends strongly on stellar type. In low-mass, solar-type stars, the wind is primarily powered by thermal pressure from a hot corona, together with magnetohydrodynamic processes providing additional acceleration. In hot, luminous massive stars however, the wind is launched by the transfer of radiative momentum to the gas via absorption and scattering in spectral lines. Because spectral lines have very large cross sections compared to continuum processes, a small fraction of the stellar luminosity suffices to accelerate substantial mass, producing terminal speeds of order a few times the stellar escape speed in massive stars. These stars therefore lose a large fraction of their mass through line-driven supersonic winds, and when two such stars form a binary their winds collide into a bow-shaped shocked region bounded by two contact discontinuities \cite{Canto_1996}. The resulting colliding-wind binary (CWB) radiates in \Halpha{}, thermal X-rays and radio, and its observational appearance encodes information about the mass-loss rates $\Mdot_{1,2}$, the terminal wind velocities $v_{\infty,1/2}$, and the orbital dynamics of the system \cite{Pejcha_2022}. Constraining these quantities is of astrophysical interest, as mass-loss shapes the evolutionary tracks and chemical yields of the most massive stars in the Universe.

In practice, inferring wind and orbital parameters from observations is challenging for two main reasons. First, the forward mapping from parameters to observables involves a three-dimensional compressible hydrodynamics simulation coupled with an N-body integrator, followed by emissivity calculations, projection, and a detector-noise model, which makes it computationally expensive. Second, the induced likelihood is implicit and cannot be evaluated pointwise, so traditional Bayesian workflows are unavailable. These two properties make the problem a natural fit for \emph{Simulation-Based Inference} \citep[SBI;][]{papamakarios_2018, greenberg_2019,Cranmer2020, sbi_package_2025}, where a neural conditional density estimator is trained on a precomputed dataset of simulation–parameter pairs and, once trained, provides amortized posterior estimates for new observations.

In this work, we demonstrate that SBI recovers the physical parameters of CWBs from short time series of synthetic \Halpha{} photon-count maps. The core ingredients are (i) a GPU-accelerated, JAX-based fluid solver \citep[astronomix;][]{storcks_astronomix_2025}, coupled with an RK4 N-body integrator for the forward model, (ii) a factorized spatio-temporal \mbox{(2+1)D} CNN architecture that is commonly used in video understanding \citep{Donahue_2015, Tran_2018} that separates spatial encoding from temporal aggregation and compresses a 10-frame $64\times64$ time series into a low-dimensional summary, and (iii) a neural spline flow \cite{durkan2019NSF} trained via amortized neural posterior estimation \cite{papamakarios_2018}. A critical observation motivates the need for time-series data in our study. The orbital elements of a CWB parametrize the binary's temporal evolution by construction, so a single \Halpha{} frame, which captures only one instant of that evolution, is structurally insufficient to constrain them. In projection, several combinations of inclination, orbital phase, eccentricity, and argument of periastron produce essentially indistinguishable 2D morphologies. The temporal evolution carries the information needed to resolve these degeneracies, including changing separation, shock-cone reorientation, and phase-dependent brightening and dimming. Spatial features simultaneously encode the wind-density structure, which motivates our spatio-temporal approach. Additionally, the classical likelihood-free alternative, Approximate Bayesian Computation \citep[ABC]{ABC_2002}, using hand-crafted summary statistics, is in principle applicable to this problem. Nevertheless, even a carefully designed physics-motivated summary fails to constrain all but the mass-loss rates, and at substantially broader posterior widths than the neural estimator, underlining the need for a learned embedding.

\begin{figure*}[t]
\centering
\includegraphics[width=1.\linewidth]{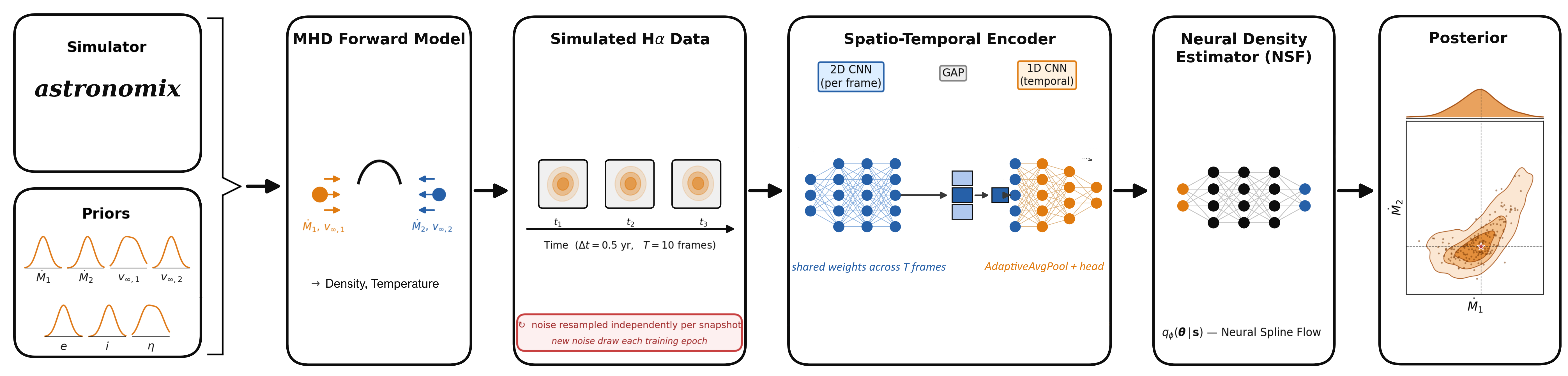}
\caption{\textbf{SBI flow chart for the CWB problem}. From a prior $p(\boldsymbol{\theta})$ over seven physical parameters, we run 3D hydrodynamic simulations with a modified version of the JAX-based fluid solver \texttt{astronomix} \cite{storcks_astronomix_2025} together with an orbit integrator and an \Halpha{} emissivity + Poisson noise pipeline to produce 10-frame photon-count time series. The resulting $\{(\boldsymbol{\theta}_i,\mathbf{x}_i)\}$ pairs train a neural spline flow with a spatio-temporal CNN embedding as the conditional density estimator, while resampling noise per simulation and per snapshot. Given a new observation, the trained estimator returns an approximate posterior $p(\boldsymbol{\theta}\,|\,\mathbf{x})$, which allows inference of stellar and orbital parameters.}
\label{fig:sbi_flow}
\end{figure*}

The goal of this study consists of three parts. \textbf{First}, we develop a complete SBI pipeline for CWBs that jointly infers seven physical parameters from short \Halpha{} time series in the presence of realistic detector noise (Fig.~\ref{fig:sbi_flow}), with posteriors verified as well-calibrated via TARP and SBC. \textbf{Second}, we show that the factorized two-stage spatio-temporal encoder design is well-matched to the physics of CWB observations, and that the temporal information resolves fundamental identifiability degeneracies that are provably unsolvable from static observations. \textbf{Third}, we demonstrate that mass-loss rates are recovered across the full prior range, wind velocities and orbital parameters are recovered when wind luminosity is sufficient, and posterior width expands appropriately in information-poor regimes.

\section{Related work}
SBI has developed rapidly in the last decade, with neural posterior, neural likelihood and neural ratio estimation now available in a common software ecosystem \cite{sbi_package_2025,papamakarios_2018, greenberg_2019, durkan2019NSF}. Applications in physics include galactic chemical evolution \citep{Guenes2025,Buck2025,Viterbo2024}, gravitational-wave astronomy \citep{Wildberger2023}, cosmology \citep[e.g.][]{Lemos2023}, galactic dynamics \citep{Viterbo2026} or supernova inference \citep{SN_SBI}. Within stellar-wind modeling, analytic thin-shell solutions provide semi-quantitative geometric insight \cite{Canto_1996}, but a data-driven, likelihood-free inference pipeline that ingests \emph{images} of CWBs has not been reported before to our knowledge. 

\section{Problem Setup and Forward Model}
\label{sec:setup}
We consider a CWB system parameterized by
\begin{equation}
  \bm{\theta} \;=\; \left(\Mdot_{1},\Mdot_{2},v_{\infty,1},v_{\infty,2},e,i,\eta\right)
  \label{eq:theta}
\end{equation}
where the first four entries are the individual mass-loss rates and terminal wind velocities, $e$ and $i$ the orbital eccentricity and inclination, and $\eta = \dot E_{\mathrm{turb}}\,/\,\langle L_{\mathrm{wind}}\rangle$
is the fraction of additional forced random turbulence  \citep[scheme adopted and modified from][]{Seo_2023} relative to the mean inserted wind luminosity. The priors, summarized in Table~\ref{tab:priors}, span ranges relevant for massive binaries and realistic geometries. The mass-loss rates and terminal wind velocities are chosen to span the parameter space of observed massive binaries.

\begin{table}[h]
  \centering
  \caption{Prior parameter distributions for our model.
  $\Mdot$ is sampled log-uniformly, $\vinf$, $e$ and $\eta$ are sampled linear-uniformly and $i$ is sampled from the standard isotropic-inclination prior.}
  \label{tab:priors}
  \small
  \setlength{\tabcolsep}{3pt}
  \begin{tabular}{@{}lll@{}}
    \toprule
    Parameter \hspace{1.5cm} & Lower \hspace{1.2cm} & Upper \hspace{1.2cm}\\
    \midrule
    $\Mdot_{1,2}$ [$\Msun\,\mathrm{yr}^{-1}$]    & $8\!\cdot\!10^{-9}$ & $1\!\cdot\!10^{-5}$ \\
    $v_{\infty,1/2}$ [$\kms$]                 & $1200$              & $3200$ \\
    $e$                                     & $0$                 & $0.85$ \\
    $i$ [deg]                               & $0$                 & $90$ \\
    $\eta$                                  & $0$                 & $0.025$ \\
    \bottomrule
  \end{tabular}
\end{table}

\subsection{Physical simulator}
The hydrodynamics are solved with a modified version of \texttt{astronomix}, a JAX-based, conservative finite-volume fluid solver
\cite{storcks_astronomix_2025,storcks2024differentiableconservativeradiallysymmetric}. The two stars are represented as point sources with an energy-and-mass wind injection scheme. The mass is deposited at a rate $\Mdot_s/V$ and the kinetic energy at a rate $\tfrac12 v_{\infty,s}^2\,\Mdot_s/V$ over a spherical injection region of volume $V$. The momentum source then arises self-consistently through the induced pressure gradient. Gravity is handled in a two-step scheme: the stars are integrated with an explicit fourth-order Runge--Kutta N-body integrator (treating the gas self-gravity as negligible compared to the stellar masses), while the gas feels the stars via a Poisson solve in which the stellar masses are deposited onto the grid with a nearest-grid-point kernel. Each simulation is run on a three-dimensional $N=64^3$ Cartesian grid for a physical time of $T_{\mathrm{end}}=5\,\mathrm{yr}$, and 10 equally spaced snapshots are stored. The simulations are initialized with a uniform ambient density of $\rho_0 = 2\,m_p\,\mathrm{cm}^{-3}$ and temperature $T_0 = 15\,000\,\mathrm{K}$, representing the warm ionized interstellar medium surrounding the binary. In this case, wind ram pressure vastly exceeds ambient thermal pressure, so shock geometry is determined by wind-wind interactions rather than external confinement. The forward problem is scale-free in $(a, \dot{M}, v_\infty)$ up to small orbital-motion asymmetries, so we fix the semi-major axis $a= 10\,\mathrm{AU}$ throughout. Additionally, we couple the sampled mass-loss rate $\Mdot$ to a stellar mass via an interpolated empirical mass / mass-loss relation $M_0(\dot{M})$ with a 5\% Gaussian spread,
\begin{equation}
  M \;\sim\; \mathcal{N}\!\left(M_0(\Mdot),\, 0.05\,M_0(\Mdot)\right).
\end{equation}
The mapping $M_0(\dot{M})$ is based on stellar evolution models from \citet{Ekstroem_2012}, from which stellar wind parameters were calculated following the procedure in \citet{Haid_2018}. The resulting distribution of orbital periods lies between $P_{\min}\approx 2.04\,\mathrm{yr}$ and $P_{\max}\approx 5.85\,\mathrm{yr}$, so that $\sim\!94\%$ of sampled systems complete at least one orbit within the observational window. This range covers the massive eccentric CWB $\eta$ Carinae \citep[$P \approx 5.54\,\mathrm{yr}$;][]{Damineli_2008}. Inference on longer-period systems is accessible to the same pipeline by retraining at an adjusted $a$ with a correspondingly extended $T_\mathrm{end}$, at the cost of additional simulation budget. Appendix~\ref{app:priors} details the empirical $M_0(\Mdot)$ relation (Fig.~\ref{fig:priors_app}) and the derived prior for the orbital period.

\subsection{From the forward simulation to synthetic observations}
\begin{figure*}[h]
  \centering
  \includegraphics[width=0.93\linewidth]{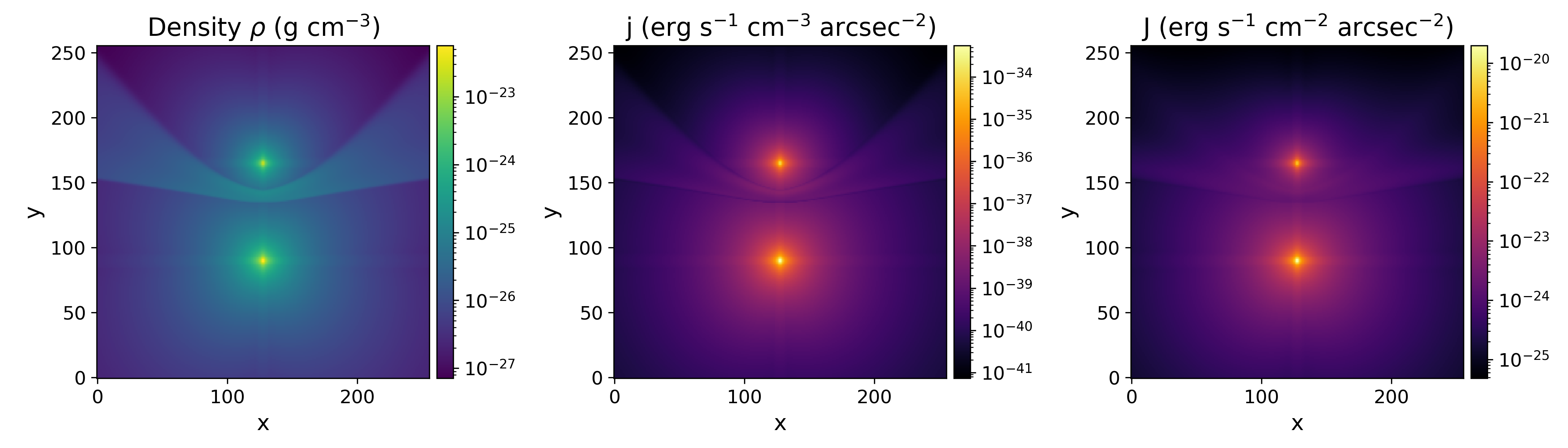}
  \caption{Forward-model pipeline for a representative CWB ($N=256^3$; $x$ and $y$ in grid units). Left to right: mass-density slice in the orbital plane; \Halpha{} emissivity $j_{\Halpha}$ (orbital plane slice); and line-of-sight-integrated \Halpha{} intensity $J$. The intensity traces the high-density bow-shock apex and the wind-collision region bounded by the two contact discontinuities.
  \label{fig:forward_model}}
\end{figure*}

Starting from the simulated density and pressure fields, the temperature is obtained from the ideal gas law, $T = \mu\,m_H P/(k_B \rho)$. Assuming a fully photoionized $\mathrm{H\ II}$ region around the hot, massive stars \citep[following the approach in][]{mackey_halpha1} we compute the H$\alpha$ emissivity from an interpolated table in \citet{osterbrock_1989}:
\begin{equation}
  j_{\Halpha}
  \;=\; 2.63\!\cdot\!10^{-33}\,\frac{n_e n_H}{T^{0.9}}
  \ \mathrm{erg\,cm^{-3}\,s^{-1}\,arcsec^{-2}},
\end{equation}
with $n_e=0.86\,\rho/m_p$ and $n_H=0.71\,\rho/m_p$. In the low-density CWB environment we adopt the optically thin limit of the radiative transfer equation, so that the specific intensity along the line of sight reduces to $J = \int j_{\Halpha}\,\mathrm{d} l$. Due to the quadratic dependence of $j_{\Halpha}$ on the density via $n_e n_H$, the emission is dominated by the bow-shock apex and the dense wind-collision region between the two contact discontinuities (see e.g. Fig.~\ref{fig:forward_model}). The intensity is then converted into an expected observed photon count by adopting representative instrument parameters (telescope diameter $D=2.4\,\mathrm{m}$, angular solid angle $A_{\mathrm{ap}}=0.04\,\mathrm{arcsec}^{2}$, exposure time $t_{\mathrm{exp}}=600\,$s, throughput $\eta_{\mathrm{tel}}=0.11$). Finally, a per-pixel noise model is applied. The most important contributions are given by the shot noise (Poisson photon-counting statistics) with variance $\sigma_{\text{Poisson}}=\sqrt{N}$, which dominates the low/medium count regime ($N\lessapprox 10^4$), and a flat-field calibration uncertainty with fractional amplitude $\epsilon_{\mathrm{flat}}=0.01$, which dominates the high count regime. The resulting dependence between physical density, emissivity and intensity for a representative simulation (without applied noise) is shown in Fig.~\ref{fig:forward_model}.

During training, each of the 10 snapshots in a simulation receives an independent noise realisation that is resampled between epochs. The network is therefore exposed to many possible observational outcomes of the same underlying physical system, which both acts as augmentation and prevents the network from memorizing a specific noise pattern.

\section{Simulation-Based Inference}
\label{sec:sbi}

We target the posterior $p(\bm{\theta}\mid \mathbf{x})$ over the seven physical parameters of Eq.~\ref{eq:theta} given an observation $\mathbf{x}\in\mathbb{R}^{T\times H\times W}$ consisting of $T=10$ noisy \Halpha{} photon-count maps of size $H=W=64$. Following the neural-posterior-estimation (NPE) paradigm \cite{papamakarios_2018}, we train a conditional density estimator $q_{\bm{\phi}}(\bm{\theta}\mid\mathbf{s})$ with parameters $\bm{\phi}$ by minimizing the expected negative log-likelihood of training parameters given the corresponding simulation summaries,
\begin{equation}
  \mathcal{L}(\bm{\phi})
  \;=\; \mathbb{E}_{p(\bm{\theta},\mathbf{x})}
    \bigl[\,-\log q_{\bm{\phi}}(\bm{\theta}\mid f_{\psi}(\mathbf{x}))\bigr],
\end{equation}
where $\mathbf{s}=f_{\psi}(\mathbf{x})$ is a neural embedding of the raw time-series cube and is trained jointly with $q_{\bm{\phi}}$. We parameterize $q_{\bm{\phi}}$ as a neural spline flow \cite{durkan2019NSF}, which we found to outperform masked autoregressive flows and mixture density networks in preliminary experiments.\footnote{The most important parts of our pipeline, including forward simulator, noise formalism, hyperparameter tuning and calibration routines, are available publicly in this GitHub repository: \href{https://github.com/n-knoell/SBI-for-CWBs}{https://github.com/n-knoell/SBI-for-CWBs}} The spline's local curvature accommodates the curved, non-linear degeneracies typical of CWB posteriors more efficiently than affine coupling blocks. All training is performed with the \texttt{sbi} python package \cite{sbi_package_2025}.

\subsection{Spatio-temporal embedding}
\label{sec:nn}
The embedding $f_{\psi}$ is a two-stage factorized spatio-temporal encoder, following approaches in video understanding \citep{Donahue_2015, Tran_2018}. The architecture has several discrete hyperparameters (written below as the sets searched over), which are tuned jointly with the flow in Sec.~\ref{sec:hyperparameter}. The selected values are reported in Table~\ref{tab:hparams}. In the first stage, a shared 2D CNN $\mathcal{E}$ of depth $L\in\{2,3\}$ with $3\!\times\!3$ strided convolutions and a doubling channel schedule $C_\ell=\min(2 C_{\ell-1}, 512)$ processes each frame independently, followed by global average pooling over the spatial axes to yield a per-frame embedding $\mathbf{e}_t$. In the second stage, the sequence $\{\mathbf{e}_t\}_{t=1}^{T}$ is passed through $L_t\in\{1,2,3\}$ 1D convolutions of kernel size $k_t=3$,
then adaptively average-pooled to $T_{\mathrm{out}}\in\{1,2,4\}$ temporal bins and mapped to the final summary by a single fully-connected head of width $d_{\mathrm{fc}}$. The overall map can be written compactly as
\begin{equation}
  f_\psi(\mathbf{x}) \;=\; h\;\circ\;\mathcal{A}\;\circ\;\mathrm{gap}\;\circ\;\mathcal{E},
\label{eq:fpsi}
\end{equation}
where $\mathcal{E}$ is the shared per-frame 2D encoder, $\mathrm{gap}$ denotes global average pooling over the spatial axes, $\mathcal{A}$ is the temporal aggregator, and $h$ is the fully-connected output head.

The choice of this embedding, rather than a single 3D CNN over the full $T \times H \times W$ cube, is motivated by how the physical parameters are encoded in the observations. First, the morphology of each individual \Halpha{} frame already carries most of the information about wind momentum ratio, bow-shock opening angle and projected inclination (see Fig.~\ref{app:geometry} for an illustration), which is a purely spatial inference task that a shared 2D encoder solves equally effectively at every time step. Second, the temporal dimension is short ($T=10$) and carries coherent global changes (orbital phase, periastron-driven brightening in eccentric orbits), not local spatio-temporal patterns. 

\emph{Sharing $\mathcal{E}$ across frames makes the per-frame feature-extraction strictly time-translation equivariant.} An observation shifted in orbital phase produces a correspondingly shifted sequence of frame embeddings, which the temporal aggregator $\mathcal{A}$ can then process consistently. \emph{This matches the physical symmetry that the orbital dynamics are invariant under a shift of the time origin}, and ensures that the network does not have to learn this invariance separately at every spatial layer.

\begin{figure*}[t]
  \centering
  \includegraphics[width=0.95\linewidth]{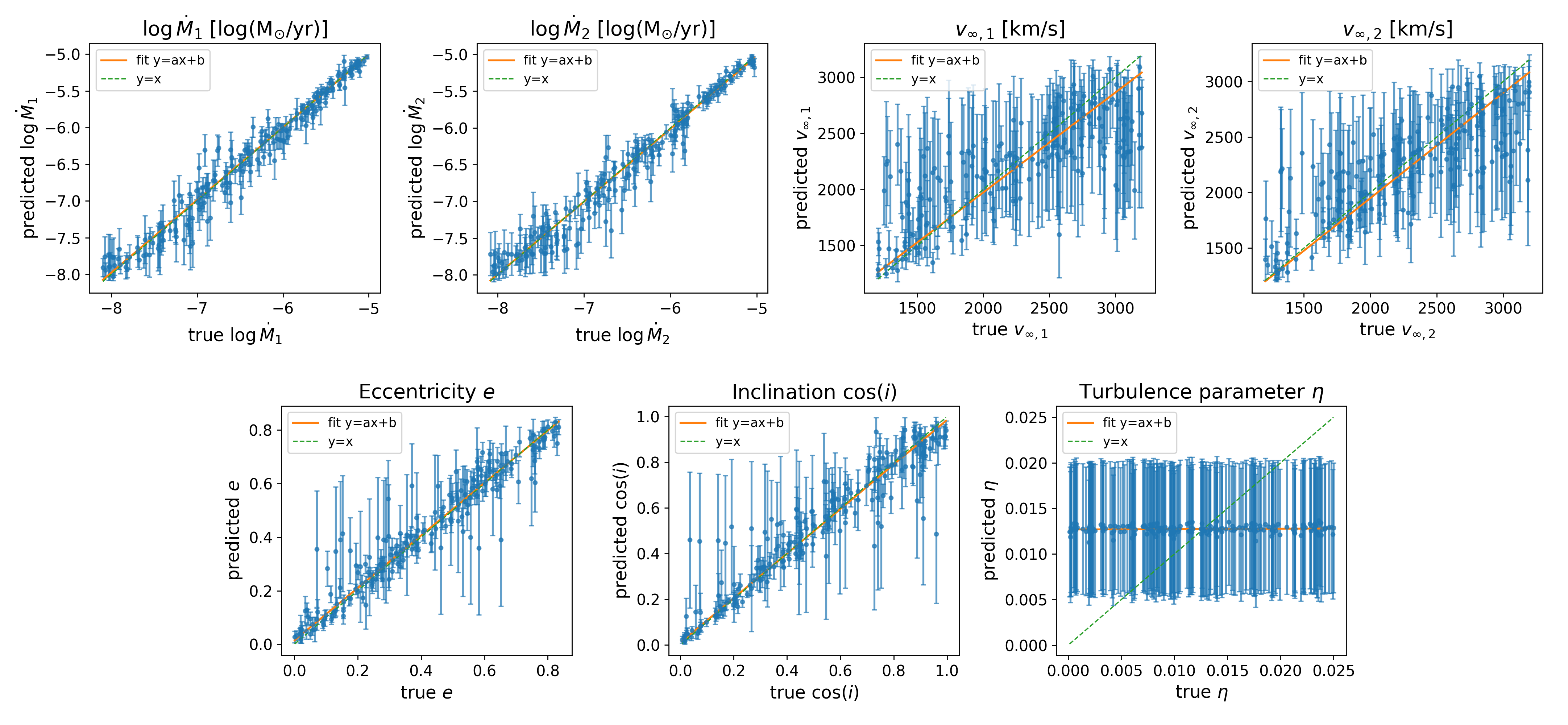}
  \caption{True vs. inferred values for a $N=200$ subset of the test simulations. Error bars are 1$\sigma$ intervals estimated from 2000 posterior samples. Mass-loss rates are recovered accurately over the 4-decade range, while terminal wind velocities slightly compress to the mean. The eccentricity $e$ and the inclination $i$ show phase-dependent recoverability. The turbulence parameter $\eta$ remains at the prior.}
  \label{fig:true_vs_inf}
\end{figure*}

\subsection{Hyperparameter search}
\label{sec:hyperparameter}
Both the embedding and the flow contain free hyperparameters. We search jointly over nine integer/categorical variables (five for the embedding CNN, two for the temporal aggregator, and two for the flow: hidden width $H_{\mathrm{flow}}\in[4,30]$ and number of transforms $N_{\mathrm{tr}}\in[3,25]$) using Optuna's NSGA-II multi-objective sampler \cite{optuna_2019, NSGAII}. Two objectives are minimized simultaneously: the validation NLL and the mean TARP deviation from the ideal coverage curve (Sec.~\ref{sec:results}). The 128-trial Pareto front optimization plot is shown in the appendix (Fig.~\ref{app:pareto}).

The final-selected configuration has $L=3$ convolutional blocks with base width $C_1=32$, a final embedding dimension $d_{\mathrm{fc}}=128$, $L_t=2$ temporal layers with stride $s_t=1$, temporal output $T_{\mathrm{out}}=2$, and a neural spline flow with $N_{\mathrm{tr}}=20$ transforms of width $H_{\mathrm{flow}}=7$ (Table~\ref{tab:hparams}). The learning rate for the Adam optimizer is $5\cdot 10^{-4}$. The flow's narrow per-transform width is a consequence of the problem's low intrinsic parameter dimensionality, which favours deeper rather than wider flows.

\begin{table}[h]
  \centering
  \caption{Best hyperparameters from the Optuna search}
  \label{tab:hparams}
  \small
  \begin{tabular}{@{}ll@{}}
    \toprule
    Hyperparameter \hspace{2.2cm} & Value \hspace{1.2cm} \\
    \midrule
    Conv blocks $L$                           & 3   \\
    Base channels $C_1$                       & 32  \\
    Final FC width $d_{\mathrm{fc}}$          & 128 \\
    Temporal layers $L_t$                     & 2   \\
    Temporal first-layer stride $s_t$         & 1   \\
    Temporal pool bins $T_{\mathrm{out}}$     & 2   \\
    Flow transforms $N_{\mathrm{tr}}$         & 20  \\
    Flow hidden features $H_{\mathrm{flow}}$  & 7   \\
    \bottomrule
  \end{tabular}
\end{table}

\section{Experiments and Results}
\label{sec:results}

\subsection{Calibration}
\label{sec:calib}
We generate $N=40,000$ simulations for training and $1,000$ additional simulations for held-out testing. On a single NVIDIA H200 GPU, we were able to run approximately $4,100$ simulations per day. The embedding and the flow are optimized jointly via amortized neural posterior estimation \cite{papamakarios_2018}.
We assess posterior calibration along two complementary axes. Tests of Accuracy with Random Points \citep[TARP;][]{tarp_2023} probes the \emph{joint} 7-D credibility coverage by constructing posterior balls around random reference points and measuring empirical containment probabilities, and perfect calibration produces a diagonal coverage curve. Simulation-based calibration \citep[SBC;][]{SBC_2018} probes the \emph{marginal} calibration by comparing the distribution of true-parameter ranks among posterior samples against the expected uniform distribution. Both diagnostics are shown in Fig.~\ref{fig:tarp_sbc_app} of the appendix. The TARP result is computed on the full $N = 1000$ held-out test set (with 1000 posterior samples per test case) with minimal coverage deviation from the optimum (Average Test Coverage $\text{ATC}=-0.061$), which indicates that the full 7-D posterior is well calibrated. On the marginal level the flow shows mild overconfidence in eccentricity and mild underconfidence in $\Mdot_2$. All other parameters are within statistical tolerance of the expected distribution under uniformity. 

\begin{figure*}[t]
  \centering
  \begin{subfigure}[b]{0.48\linewidth}
    \includegraphics[width=\linewidth]{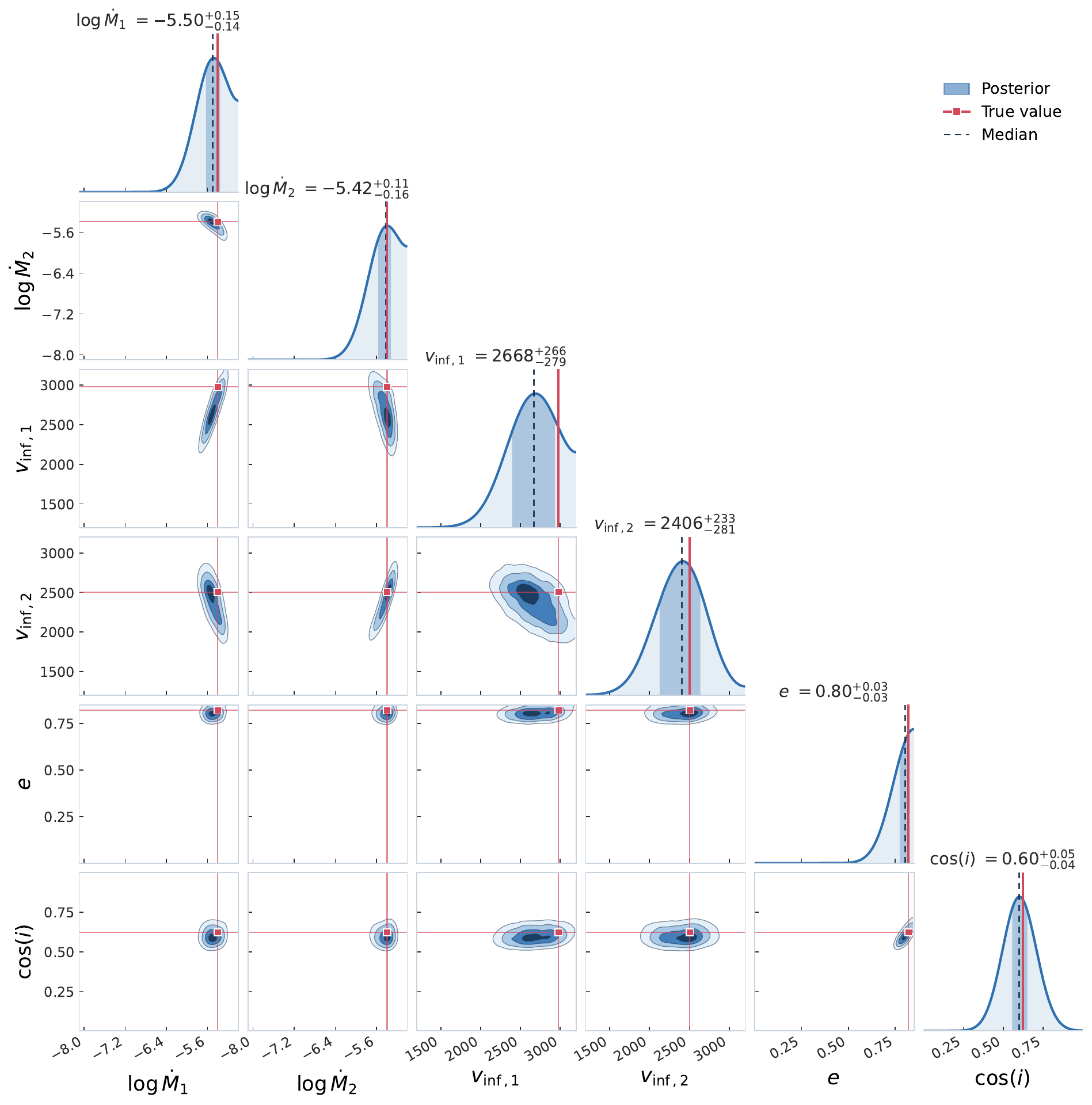}
    \caption{Case~1 (high wind luminosity, low relative noise).}
  \end{subfigure}\hfill
  \begin{subfigure}[b]{0.48\linewidth}
    \includegraphics[width=\linewidth]{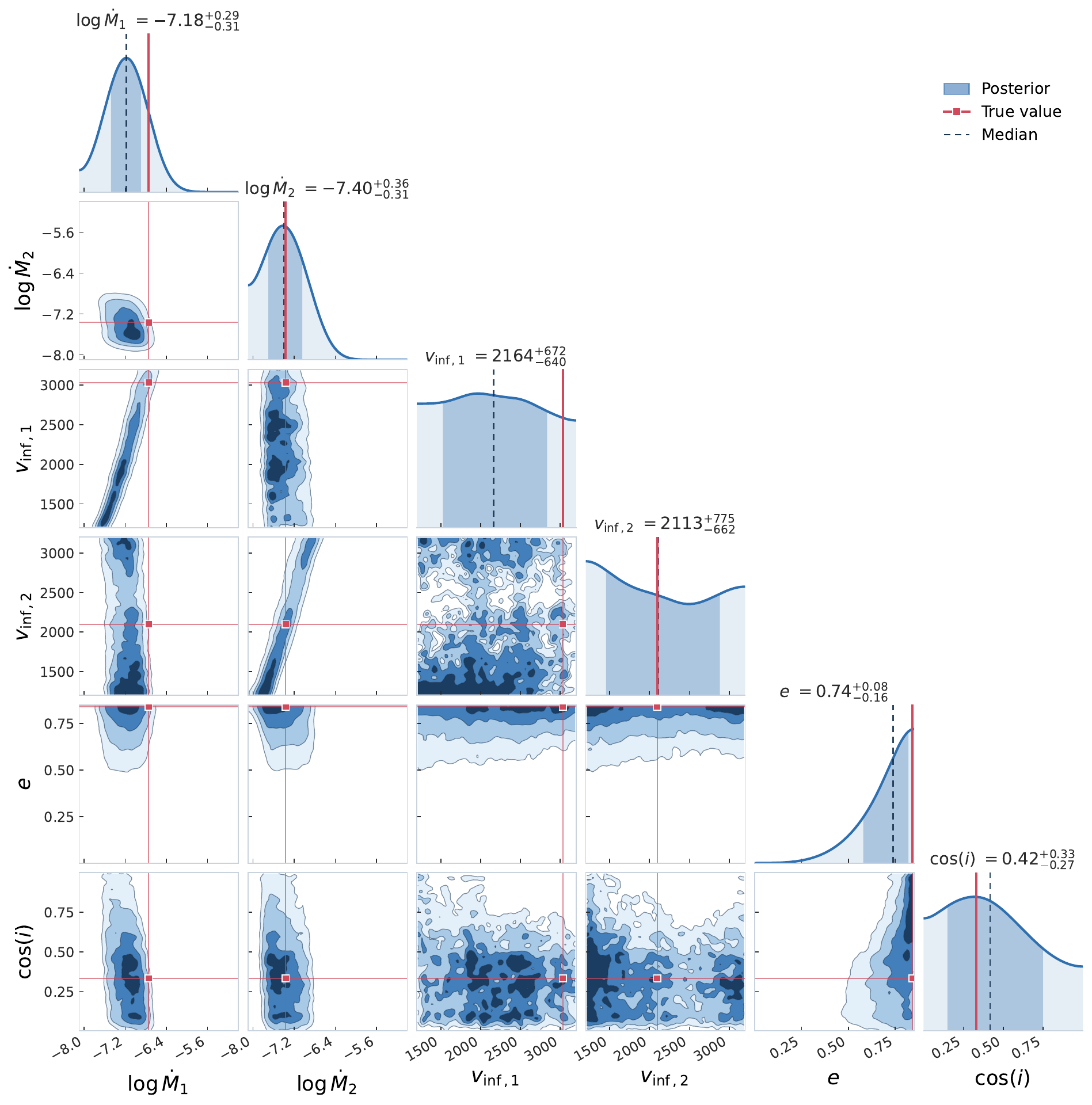}
    \caption{Case~2 (low wind luminosity, high relative noise).}
  \end{subfigure}
  \caption{Inferred posterior distributions and parameter correlations for two reference CWBs. Diagonal panels show the one-dimensional marginal posteriors with the shaded band marking the central 68\% credible interval and the dashed line the posterior median. Off-diagonal panels show the two-dimensional joint posteriors as filled contours enclosing the $0.5\sigma$, $1\sigma$, $1.5\sigma$ and $2\sigma$ credible regions. Solid red lines indicate the true parameter values. In the high-luminosity regime (a) all parameters (except $\eta$) are tightly constrained. In the low-luminosity regime (b) the shock features are washed out by Poisson noise, essentially all information about the terminal wind velocities is lost, and the corresponding marginals approximately return the prior.}
  \label{fig:posteriors_ref}
\end{figure*}

\subsection{Point estimates and posteriors}

Figure~\ref{fig:true_vs_inf} compares the posterior mean and $1\sigma$ interval against the true parameters for $200$ randomly chosen test simulations. The mass-loss rates are recovered accurately over the four decades of prior range. Terminal wind velocities are systematically recovered but with larger per-simulation error bars and a mild tendency to under-predict at the high end of the prior. Eccentricity and inclination show a bimodal error pattern, with many simulations pinned to narrow posteriors and others retaining significant prior width, reflecting whether the 5-year observational window happens to capture dynamically informative portions of the orbit (particularly the periastron passage). The turbulence parameter $\eta$ is essentially unconstrained, as the posterior is statistically indistinguishable from the prior across the full range. We believe this is caused by a combination of (i) the artificial detector noise dominating the much weaker turbulent signal and (ii) strided pooling in the embedding filtering out small-scale turbulent texture. We will return to this point in Sec.~\ref{sec:discussion}.

\paragraph{Reference cases.}

Table~\ref{tab:cases} lists three representative held-out simulations spanning different wind-luminosity regimes. Corner plots of the inferred posteriors for Case~1 (high luminosity) and Case~2 (low luminosity) are shown in Fig.~\ref{fig:posteriors_ref}. Case~3 (highly asymmetric wind strengths) is in the appendix (Fig.~\ref{fig:posterior_mixed_app}) together with the full time series of Case~1 (Fig.~\ref{fig:ref_high_app}). It should be noted that the turbulence parameter $\eta$ is omitted here, because it resembles the prior in all cases, as already mentioned throughout this work.

\begin{table}[t]
  \centering
  \caption{Three reference cases spanning wind-luminosity regimes.}
  \label{tab:cases}
  \small
  \setlength{\tabcolsep}{3pt}
  \renewcommand{\arraystretch}{1.2}
  \begin{tabular}{@{}cccccccc@{}}
    \toprule 
    \# & $\Mdot_1$ & $\Mdot_2$ & $v_{\infty,1}$ & $v_{\infty,2}$ & $e$ & $\cos i$ & $\eta$ \\
    \midrule
    1 & $4\!\cdot\!10^{-6}$ & $4\!\cdot\!10^{-6}$ & $2977$ & $2505$ & $0.82$ & $0.62$ & $0.016$ \\
    2 & $1.8\!\cdot\!10^{-7}$ & $4.3\!\cdot\!10^{-8}$ & $3033$ & $2097$ & $0.84$ & $0.33$ & $0.011$ \\
    3 & $8.4\!\cdot\!10^{-6}$ & $1.2\!\cdot\!10^{-8}$ & $3080$ & $1740$ & $0.09$ & $0.89$ & $0.007$ \\
    \bottomrule
  \end{tabular}
\end{table}

\begin{figure*}[t]
  \centering
  \includegraphics[width=0.9\linewidth]{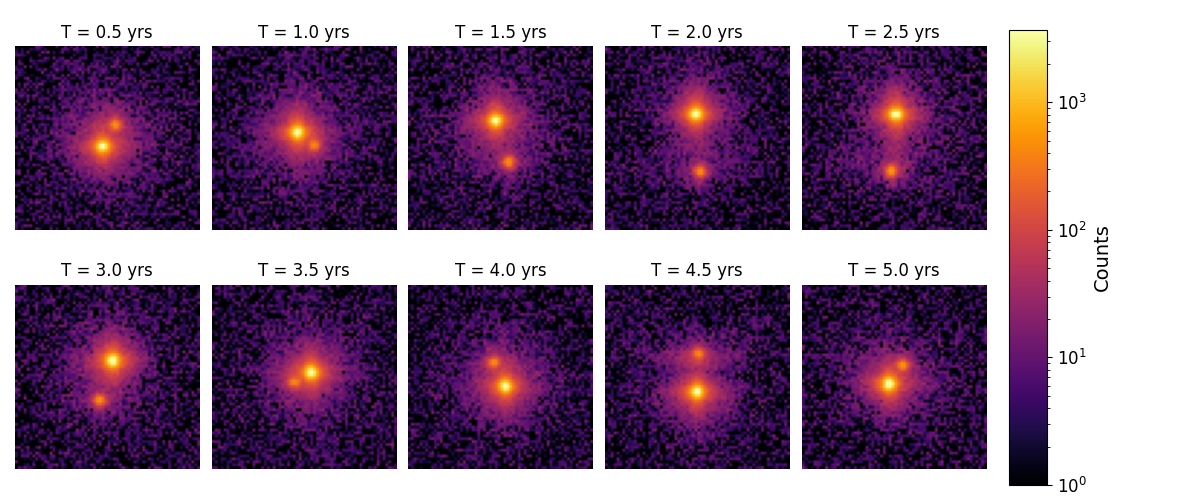}
  \caption{Complete 10-frame \Halpha{} photon-count time series for Case~2 (low wind luminosity, high relative noise). Noise dominates in most snapshots and shock features and the stagnation point are only intermittently visible.}
  \label{fig:ref_low_app}
\end{figure*}

In Case~1 both stars generate strong winds and the shocked collision region is dense and bright. As a result, the fractional Poisson noise per-pixel is very low, allowing all parameters except $\eta$ to be tightly constrained around their true values (see left panel of Fig.~\ref{fig:posteriors_ref}). In Case~2 the wind luminosities of both stars are very low (the full time series for Case~2 is shown in Fig.~\ref{fig:ref_low_app}). Consequently, noise dominates in most snapshots and the stagnation point along with much of the shock structure is only intermittently visible. Mass-loss rates and orbital parameters $e$ and $i$ are still reasonably well estimated (see right panel of Fig.~\ref{fig:posteriors_ref}), because they are encoded in the overall brightness and large-scale geometry. However, the terminal-velocity posteriors approximately coincide with the prior, because the distinctive features that encode $v_\infty$ (opening angle, width of the shocked region, and periastron asymmetry) are lost in the noise and most of the shock structure is washed out (see Fig.~\ref{fig:ref_low_app}). Case~3 is a highly asymmetric case in which one strong wind dominates. Now, the recovery pattern matches expectations, namely that $v_{\infty,1}$ is well constrained while $v_{\infty,2}$ returns approximately the prior. In addition, eccentricity and inclination are still very well constrained (see Fig.~\ref{fig:posterior_mixed_app} in the appendix).

\paragraph{Effect of training-set size.}
We retrain the selected architecture on seven subsets $N\in\{10^2,5\!\cdot\!10^2,10^3,5\!\cdot\!10^3,10^4,2\!\cdot\!10^4,4\!\cdot\!10^4\}$ of the full set of training simulations to map out how posterior quality scales with the available simulation budget. The resulting RMSE and posterior-width curves are provided in Appendix~\ref{app:trainsize} (Fig.~\ref{fig:set_size_app}). Both metrics remain essentially flat until $N\approx10^4$ and then improve rapidly until  $N\approx2\cdot10^4$. Further small improvement is still observed out to our largest size $N=4\cdot10^4$ without fully plateauing, suggesting that additional training data would still slightly improve RMSE and tighten the posteriors. However, pushing $N$ further is costly, as the $\sim\!4{,}100$ sims/day throughput reported above means that an additional $4\!\cdot\!10^4$ simulations would consume roughly 10 GPU-days on a single NVIDIA H200, and we therefore report results at the largest size afforded by our computational budget. The turbulence parameter $\eta$ is the only exception to this scaling pattern, remaining uninformative at every value of $N$ tested.

\paragraph{Baseline comparison (ABC).}
\phantomsection\label{sec:abc}
To assess whether the learned spatio-temporal embedding is actually needed, or whether a carefully designed low-dimensional summary would already suffice, we compare against an Approximate Bayesian Computation (ABC) rejection sampler operating on a 62-dimensional hand-crafted summary of the same 10-frame \Halpha{} time series. The summary stacks per-frame log-total flux and log-peak flux (to target $\Mdot$ through $n_e n_H$), brightness-weighted centroids (to capture orbital motion) and second moments (bow-shock extent and, therefore, $v_\infty$), together with two global temporal descriptors that test for phase-driven variability (due to the eccentricity). The full definitions and results of this process are given in Appendix~\ref{app:abc}. Figure~\ref{fig:abc_posteriors_app} shows the resulting ABC posteriors for the same reference high- and low-luminosity cases as in Fig.~\ref{fig:posteriors_ref}. In both cases ABC recovers only the mass-loss rates, and at substantially broader widths than the neural estimator. Additionally, $v_{\infty,1,2}$, $e$, $\cos i$ and $\eta$ effectively return the prior. This is consistent with $\Mdot$ dominating the integrated \Halpha{} flux, while the more subtle morphological and temporal signatures that encode the orbit and wind kinematics are destroyed when $T\!\times\! H\!\times\! W = 40{,}960$ pixels are collapsed to $62$ scalars. In other words, classical summary-statistic ABC does not scale to the dimensionality of the CWB problem, and the NPE approach is a necessary rather than optional component of our pipeline.

\section{Discussion and Limitations}
\label{sec:discussion}
Three empirical patterns emerge from our experiments. First, mass-loss rates are by far the easiest parameter group, because they modulate the overall brightness of the bow-shock quadratically through $n_e n_H$, and this dominant signal survives even in the low-luminosity regime. Second, terminal wind velocities are encoded in higher-order morphological features (bow-shock opening angle, the width of the shocked layer, and the left--right asymmetry induced by orbital motion), which are precisely the features that are first erased by Poisson noise. Third, the turbulence parameter $\eta$ is never informative. A possible explanation is that this results from a combination of the artificial per-pixel noise with amplitude comparable to the turbulent signal, and the information-bottleneck character of the spatial-pooling layers, which average out the small-scale turbulent texture that would be needed to determine $\eta$. A plausible mitigation would be an embedding that explicitly retains high-frequency spatial content, for example through a parallel branch operating on spectral or wavelet features \citep[e.g.][]{Rost2025}. We note that $\eta$ was introduced primarily to inject small-scale turbulence into the observations, not as a target of scientific interest. Its non-recovery therefore does not affect the parameters that matter physically (mass-loss rates, wind velocities, and orbital elements).

The ABC baseline shows that even a deliberately rich $62$-D hand-crafted summary, built to explicitly target the physical signals in the data, recovers only mass-loss rates (with substantially broader posteriors) and leaves orbital and kinematic parameters at the prior. The learned spatio-temporal embedding therefore enables recovery of parameters that classical methods leave essentially unconstrained.

\paragraph{Model misspecification.}
All experiments use synthetic data generated by the same simulator that trains the network. A real-world deployment would require addressing the inevitable simulator-observation gap: stellar radiation fields, dust absorption, instrumental point-spread functions, and intrinsic stellar variability are all absent from our forward model. Recent work on robust NPE and misspecified simulators has introduced a range of strategies to identify model misspecification and perform amortized model selection \citep[e.g.][]{Buerkner2022,radev2023,Guenes2025}. 

\section{Conclusion}
We have presented an amortized simulation-based inference framework that recovers the physical parameters of colliding-wind binaries from short, noisy \Halpha{} time series, yielding well-calibrated posteriors in a challenging seven-dimensional inverse problem. Our findings point to a broader principle, namely that in many simulator-based settings non-identifiability arises not from model limitations but from insufficient observations. We show that even short time series can break these degeneracies, while single-frame observations remain fundamentally ambiguous. Crucially, this requires architectures that reflect the structure of the data. By adopting a factorized spatial-then-temporal encoder, our model aligns with the underlying physics and improves identifiability compared to unstructured approaches. We further observe that the learned posterior adapts to the information content of the data, tightening in high-signal regimes and reverting to the prior when observations are dominated by noise. This behavior is essential for reliable uncertainty quantification in likelihood-free inference and highlights a key advantage of amortized neural methods over classical approaches. Looking forward, bridging the gap to real observations and integrating multi-modal data will be critical next steps. More generally, our results suggest that structured amortization, which combines inductive biases with flexible density estimators, offers a scalable path toward reliable inference in complex, high-dimensional scientific models.

\section*{Impact Statement}
This paper presents work that aims to advance the field of machine learning for the physical sciences. We do not foresee direct societal risks or applications beyond the scientific community, and the methods developed here are generic tools for likelihood-free inference on simulator-based forward models.

\section*{Acknowledgements}
This work is funded by the Carl-Zeiss-Stiftung through the NEXUS program. This work was supported by the Deutsche Forschungsgemeinschaft (DFG, German Research Foundation) under Germany’s Excellence Strategy EXC 2181/1 - 390900948 (the Heidelberg STRUCTURES Excellence Cluster). We acknowledge the usage of the AI-clusters Tom and Jerry funded by the Field of Focus 2 of Heidelberg University.


\bibliography{papers}
\bibliographystyle{icml2026}

\newpage
\appendix
\onecolumn

\section{Prior distributions}
\label{app:priors}

\begin{figure}[h]
  \centering
  \begin{subfigure}[b]{0.49\linewidth}
    \includegraphics[width=\linewidth]{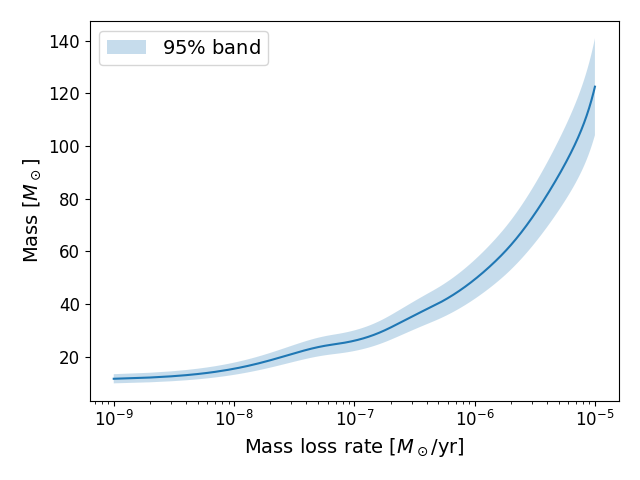}
    \caption{$M_0(\Mdot)$ with 5\% Gaussian scatter.}
  \end{subfigure}\hfill
  \begin{subfigure}[b]{0.49\linewidth}
    \includegraphics[width=\linewidth]{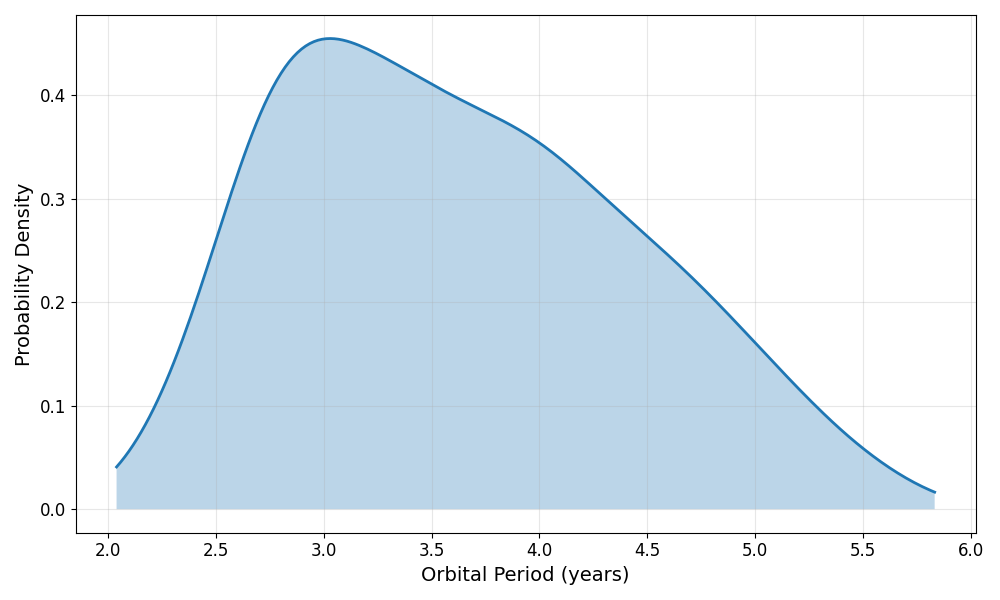}
    \caption{Prior-induced period distribution.}
  \end{subfigure}
  \caption{Derived quantities from the mass-loss prior. (a) the empirical mass / mass-loss mapping with Gaussian scatter used to draw stellar masses, and (b) the resulting prior distribution of orbital periods.}
  \label{fig:priors_app}
\end{figure}

Figure~\ref{fig:priors_app} shows the empirical mass / mass-loss relation $M_0(\Mdot)$ used to assign a stellar mass to each sampled $\Mdot$ (with $5\%$ Gaussian scatter), together with the prior-induced distribution of orbital periods for a binary with semi-major axis $a=10\,\mathrm{AU}$.

\section{Geometry of CWBs}
\label{sec:geometry}
\begin{figure}[h]
\centering
\includegraphics[width=0.5\linewidth]{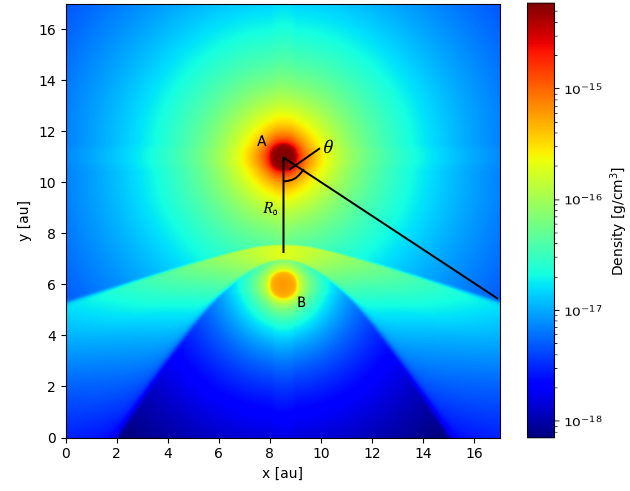}
\caption{Mass-density slice in the orbital plane of a stationary CWB, showing the stand-off distance $R_{0}$ measured from the stronger-wind star~A (Eq.~\ref{eq:R0}) and the opening angle $\theta$ of the bow-shock cone (becomes $\theta_\infty$ for $r\rightarrow\infty$;  Eq.~\ref{eq:theta_inf}). Example parameters:
$\dot{M}_{1}=5.2\cdot 10^{-6}~\text{M}_\odot/\text{yr}$,
$\dot{M}_{2}=5.2\cdot 10^{-7}~\text{M}_\odot/\text{yr}$,
$v_{\infty,1}=2200~\text{km}/\text{s}$,
$v_{\infty,2}=2000~\text{km}/\text{s}$,
$a=5~\text{AU}$}
\label{app:geometry}
\end{figure}

For physical intuition, it is helpful to introduce the analytical thin-shell solution of \citet{Canto_1996} for two interacting, axisymmetric, hypersonic, non-accelerating winds with efficient radiative cooling, launched from a stationary binary (orbital motion neglected).
The resulting geometric properties and surface-density distributions provide useful insight, even when these assumptions are only approximately satisfied. For two stars with masses $M_1, M_2$ and wind velocities $v_{\infty,1}, v_{\infty,2}$ it is useful to start by defining the ratios of momenta $\beta$ and terminal wind velocities $\alpha$ as
\begin{equation}
    \alpha \;\equiv\; \frac{v_{\infty,1}}{v_{\infty,2}},
    \qquad
    \beta \;\equiv\; \alpha\,\frac{\dot{M}_{1}}{\dot{M}_{2}}
\end{equation}
Under the above listed assumptions of constant-velocity, isotropic winds, mass conservation yields the individual density profiles
\begin{equation}
    \rho_1(r)=\frac{\dot M_1}{4\pi r^2 v_{\infty,1}}, \quad \rho_2(r_2)=\frac{\dot M_2}{4\pi r_2^2 v_{\infty,2}}.
\end{equation}
where $r$ and $r_2$ denote the distances from star 1 and star 2, respectively. In this case, the stand-off distance (stagnation radius), which is the point along the connection line of both stars where the ram pressures $P=\rho v_\infty^2$ are in balance, is given by
\begin{equation}
\label{eq:R0}
    R_{0}=a \frac{\sqrt{\beta}}{1+\sqrt{\beta}}
\end{equation}
with orbital separation $a$. For $\beta > 1$ we have $R_{0} > a/2$, i.e.\ the stagnation point lies closer to the weaker-wind star, as expected. By combining mass and momentum conservation with geometric arguments, one obtains an expression for the asymptotic opening angle $\theta_\infty$ of the bow-shock cone, measured at star 1 from the inter-stellar axis to the asymptotic shell direction, in the limit $r\rightarrow \infty$:
\begin{equation}
\label{eq:theta_inf}
    \theta_\infty-\tan \theta_\infty=\frac{\pi}{1-\beta}.
\end{equation}
For an illustration of both quantities, see Fig.~\ref{app:geometry}.

\section{Hyperparameter tuning}
\begin{figure}[h]
  \centering
  \includegraphics[width=0.5\linewidth]{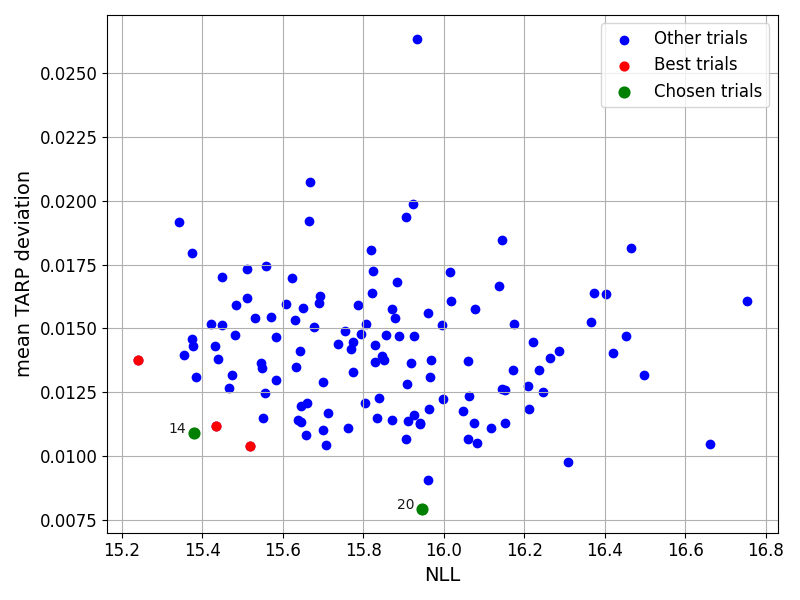}
  \caption{Pareto front over 128 Optuna trials for the joint minimization of validation NLL and mean TARP deviation. Red points mark the Pareto-optimal set. Trials 14 and 20 (green) were retrained at full batch, trial~20 giving the best final calibration.}
  \label{app:pareto}
\end{figure}

Fig.~\ref{app:pareto} shows the 128-trial Pareto front optimization plot, in which the validation NLL and the mean TARP deviation from the ideal coverage curve are optimized simultaneously. The green marked and annotated trial~20 giving the best final calibration among the chosen best trials for a  training on the complete test simulation set ($N=40,000$). 

\section{Effect of training-set size}
\label{app:trainsize}
\begin{figure}[h]
  \centering
  \begin{subfigure}[b]{0.49\linewidth}
    \includegraphics[width=\linewidth]{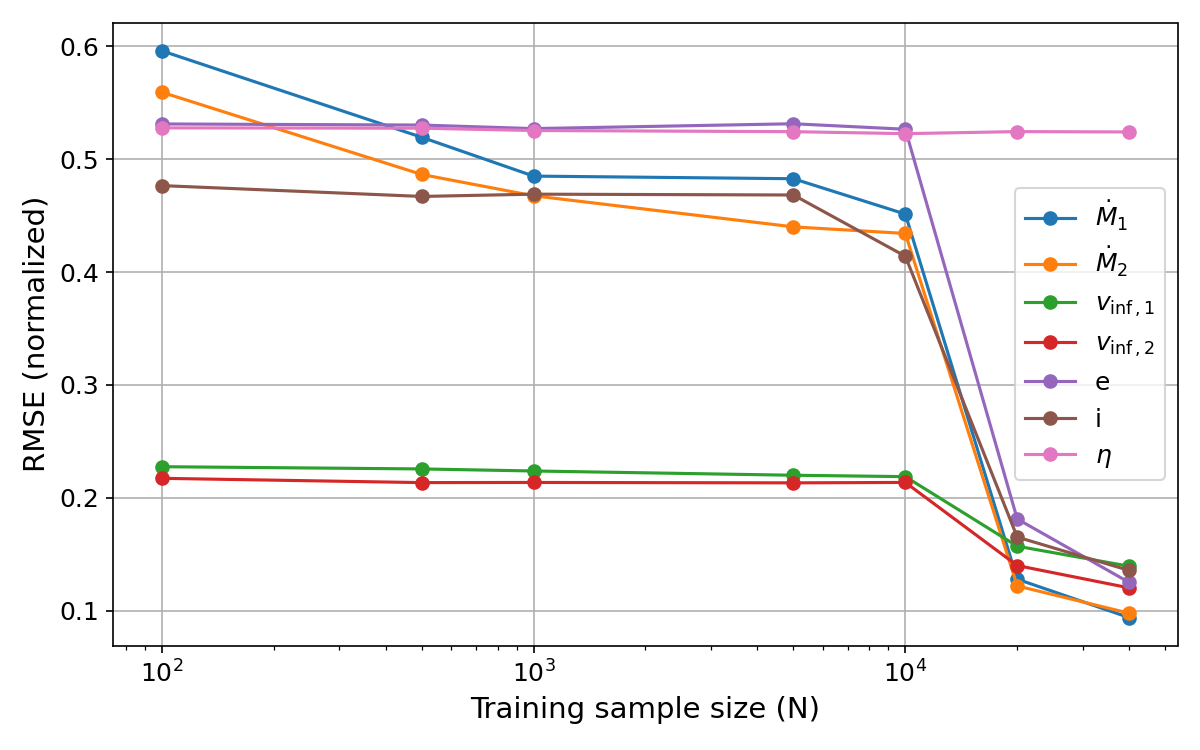}
    \caption{RMSE vs.\ true values.}
  \end{subfigure}\hfill
  \begin{subfigure}[b]{0.49\linewidth}
    \includegraphics[width=\linewidth]{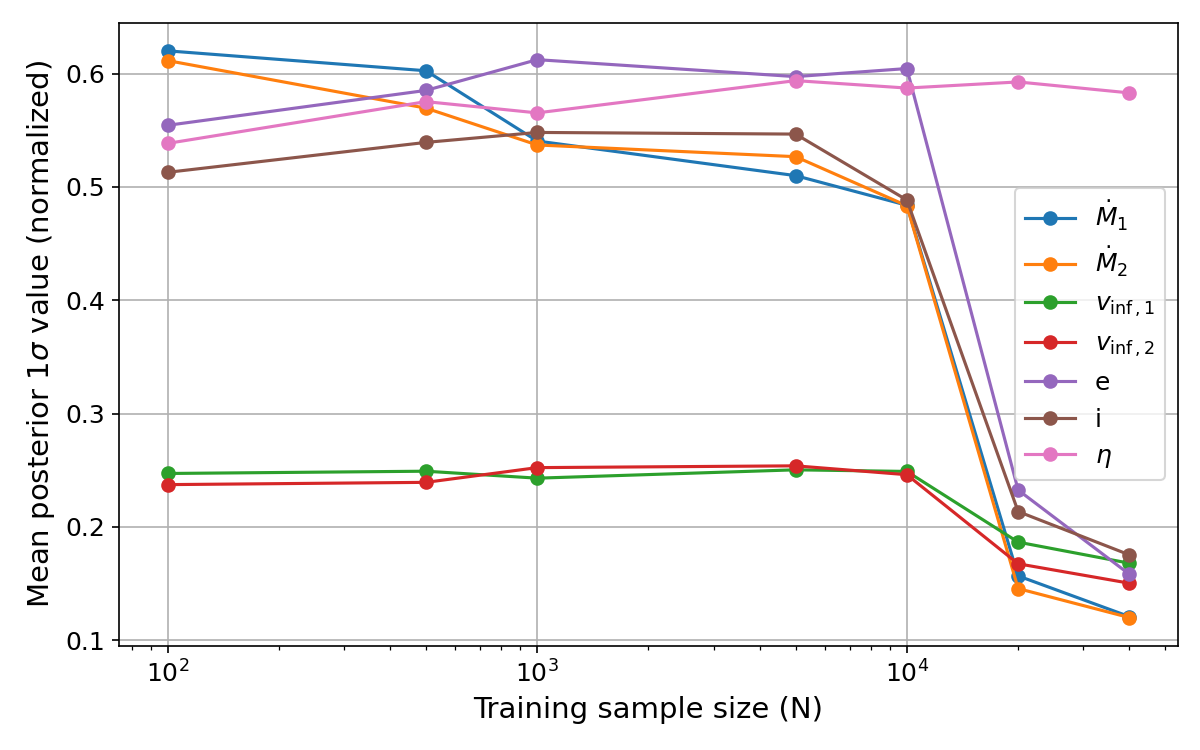}
    \caption{Posterior $1\sigma$ width.}
  \end{subfigure}
  \caption{RMSE and posterior width as a function of training-set size. Both metrics saturate slowly past $N\approx 2\cdot 10^4$ simulations.}
  \label{fig:set_size_app}
\end{figure}

Figure~\ref{fig:set_size_app} shows how the quality of inferred posteriors depends on the number of training simulations. We retrain the selected architecture on the subsets $N\in\{10^2, 5\!\cdot\!10^2,10^3,5\!\cdot\!10^3,10^4,2\!\cdot\!10^4,4\!\cdot\!10^4\}$ of the training simulation set and evaluate on the held-out test set. Both the RMSE of the posterior mean against the ground truth and the $1\sigma$ Gaussian-approximated posterior width are approximately flat up to $N \approx 10^4$, then improve rapidly until $N \approx 2\cdot 10^4$, and slowly flatten with only a small residual gain out to the largest training size. This suggests that additional simulations would further reduce bias and sharpen posteriors, but with diminishing returns. The turbulence parameter $\eta$ is an exception, as it remains uninformative across all training-set sizes, consistent with the discussion in Sec.~\ref{sec:discussion}. Note that the hyperparameters were held fixed at the values of Table~\ref{tab:hparams} (tuned at the largest $N$) for all training-set sizes. Per-$N$ retuning might improve the smaller-$N$ results slightly, which makes the curves a conservative estimate of the achievable performance at small sample sizes.

\section{Calibration diagnostics}
\label{app:calib}

\begin{figure}[h]
  \centering
  \begin{subfigure}[t]{0.48\linewidth}
    \includegraphics[width=\linewidth]{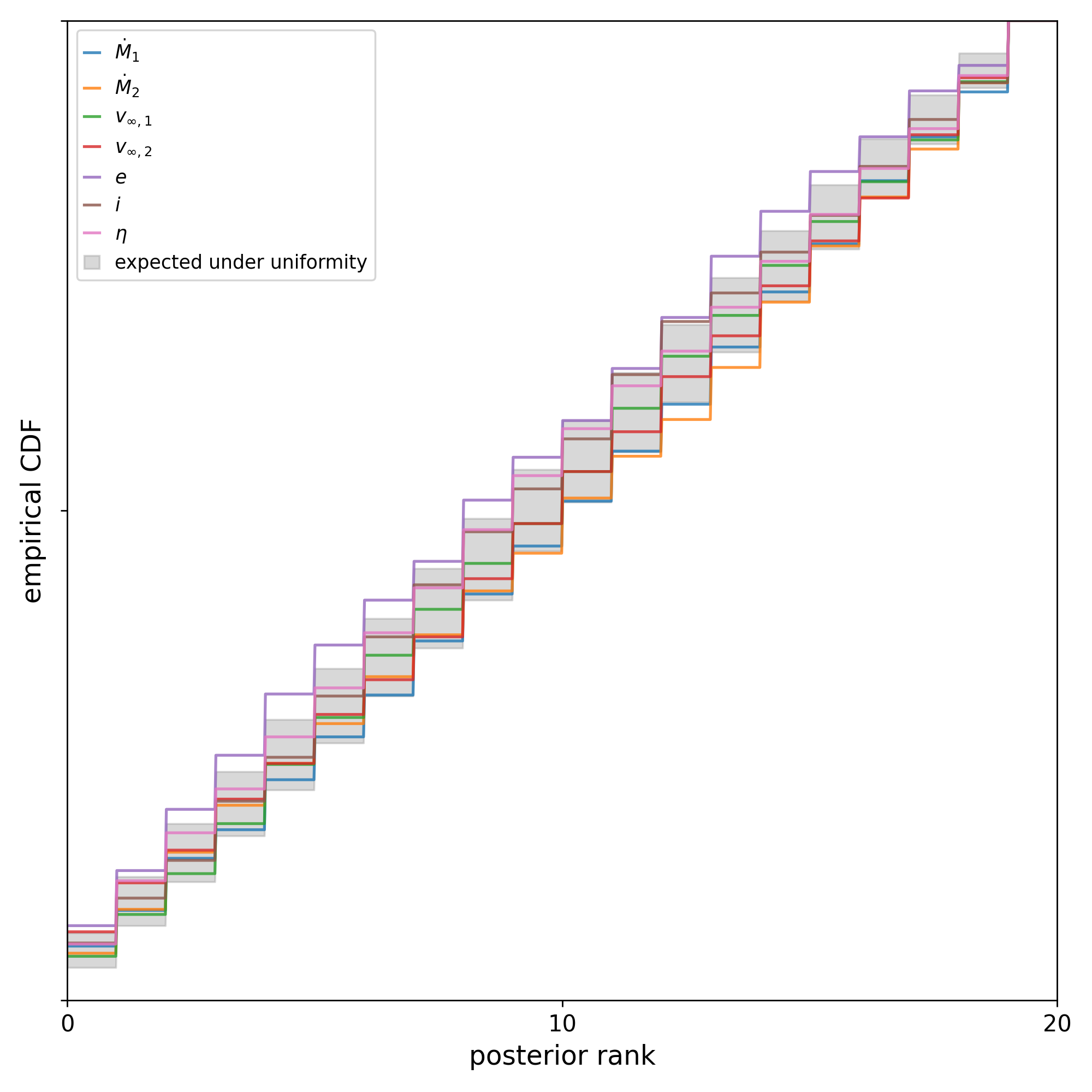}
    \caption{SBC rank plot, per parameter.}
  \end{subfigure}\hfill
  \begin{subfigure}[t]{0.492\linewidth}
    \includegraphics[width=\linewidth]{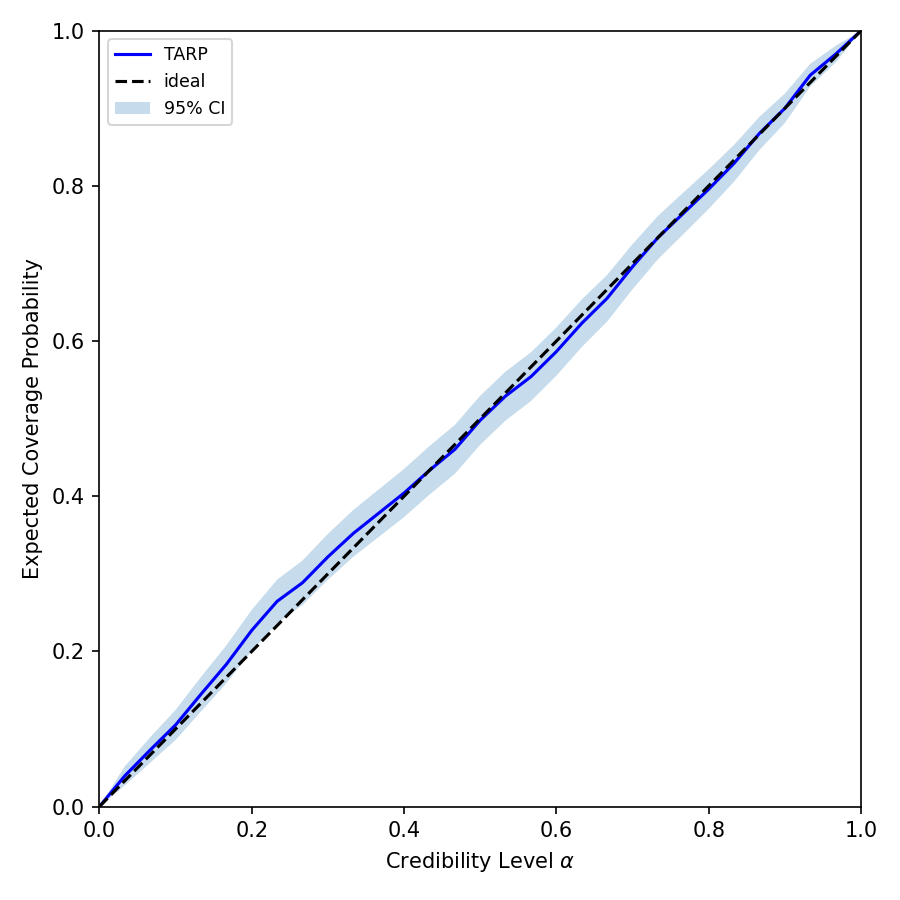}
    \caption{TARP joint coverage.}
  \end{subfigure}
  \caption{Posterior calibration diagnostics for the 7-parameter model. (a) simulation-based calibration rank plot are within  uniform. (b) the TARP coverage curve tracks the diagonal across all credibility levels $\alpha$.}
  \label{fig:tarp_sbc_app}
\end{figure}

Figure~\ref{fig:tarp_sbc_app} shows the joint TARP coverage curve and the per-parameter SBC plot for the trained model. The TARP curve tracks the ideal diagonal across all credibility levels, indicating that the full 7-dimensional posterior is jointly well-calibrated ($\text{ATC}=-0.061$). The SBC as an empirical CDF vs. posterior rank plot confirms an approximately uniform rank distribution for mass-loss rate~1, both terminal wind velocities, inclination, and the turbulence parameter. Slight deviations from the optimum band indicate mild overconfidence for eccentricity ($-8.8\%$ too narrow posteriors) and mild underconfidence ($5.1\%$ too wide posteriors) for $\Mdot_2$. 

\section{Additional reference cases}
\label{app:extra}

\begin{figure}[h]
  \centering
  \includegraphics[width=0.8\linewidth]{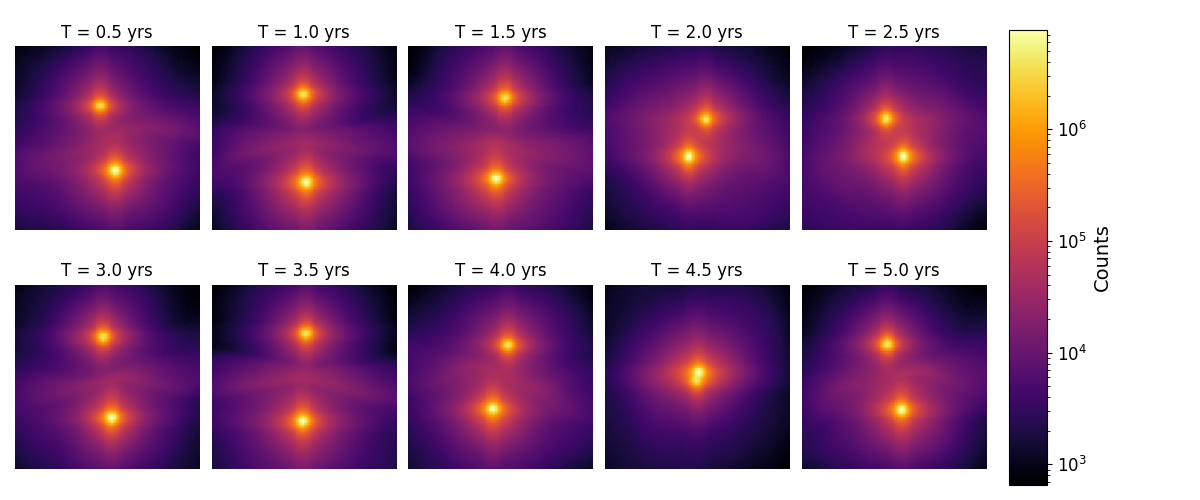}
  \caption{Complete 10-frame \Halpha{} photon-count time series for
  Case~1 (high wind luminosity, low relative noise, highly eccentric orbit). The periastron passage produces a clear brightening and reshaping of the bow shock.}
  \label{fig:ref_high_app}
\end{figure}

Figure~\ref{fig:ref_high_app} shows the full 10-frame time series for Case~1 of Table~\ref{tab:cases}, the high-luminosity counterpart of Case~2 above. The bow shock, the two contact discontinuities, and the periastron-driven brightening are all visible at a per-pixel relative noise of a few percent.

\begin{figure}[h]
  \centering
  \includegraphics[width=0.54\linewidth]{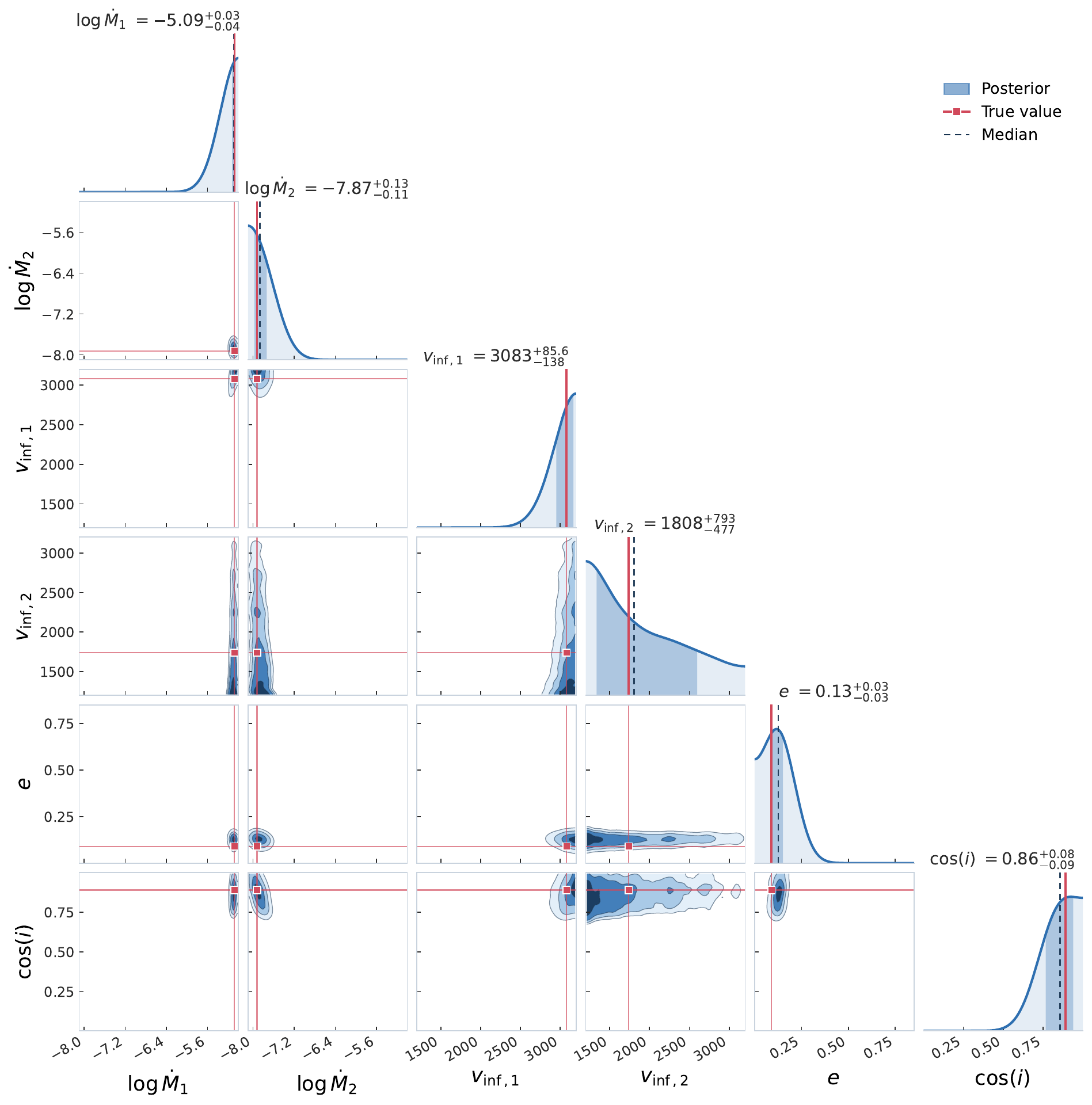}
  \caption{Inferred posterior for Case~3, a highly asymmetric CWB with one dominant wind. The dominant star's $\vinf$ is well constrained, the secondary's $\vinf$ returns a slightly distorted prior.}
  \label{fig:posterior_mixed_app}
\end{figure}

Figure~\ref{fig:posterior_mixed_app} shows the full posterior corner plot for Case~3 of Table~\ref{tab:cases}: a strongly asymmetric CWB with
$\Mdot_1/\Mdot_2\sim 7\!\cdot\!10^{2}$. As anticipated, the dominant wind's velocity $v_{\infty,1}$ is well constrained, while $v_{\infty_,2}$ returns a slightly skewed prior, because the weaker wind's shock features are no longer above the noise floor. Both mass-loss rates and orbital parameters $e$ and $i$ remain robustly recovered.

\section{ABC baseline: summary statistics and posteriors}
\label{app:abc}

As a classical likelihood-free baseline we implement an Approximate Bayesian Computation (ABC) rejection sampler on a physics-motivated summary of the same 10-frame \Halpha{} photon-count time series that our neural pipeline takes in. This appendix defines the summary statistic, the distance, the acceptance threshold, and shows the resulting posteriors for the two reference cases of Sec.~\ref{sec:results}.

\subsection{Summary statistic}
\label{app:abc_summary}
For an observation $I$ of the \Halpha{} photon count with $T=10$ time frames and $H=W=64$ pixels per spatial dimension we define a $(6T+2)=62$-dimensional summary vector
\begin{equation}
  s(\mathbf{I}) \;=\;
  \bigl\{\,
    \log_{10} F_t,\;
    \log_{10} P_t,\;
    c_{x,t},\;
    c_{y,t},\;
    \sigma_{x,t},\;
    \sigma_{y,t}
  \bigr\}_{t=1}^{T}
  \;\cup\;
  \{\phi,\ \pi\}
  \label{eq:abc_summary}
\end{equation}
with per-frame components (and using $p,q$ as pixel row/column indices). $F_t$ is the total flux and $P_t$ the peak flux, $ c_{x,t}$ and $c_{y,t}$ are brightness-weighted centroids and second moments $\sigma_{x/y,t}^{2}$, 
\begin{align}
  F_{t}   &= \sum_{p,q} I_{t,p,q}, & P_{t}   &= \max_{p,q} I_{t,p,q},\\
  c_{x,t} &= \frac{1}{F_{t}}\sum_{p,q} p\,I_{t,p,q}, &
  c_{y,t} &= \frac{1}{F_{t}}\sum_{p,q} q\,I_{t,p,q},\\
  \sigma_{x,t}^{2} &= \frac{1}{F_{t}}\sum_{p,q}(p-c_{x,t})^{2}\,I_{t,p,q}, &
  \sigma_{y,t}^{2} &= \frac{1}{F_{t}}\sum_{p,q}(q-c_{y,t})^{2}\,I_{t,p,q},
\end{align}
together with two global temporal descriptors 
\begin{equation}
\phi \;=\; \frac{\operatorname{std}_{t}(F_{t})}{\operatorname{mean}_{t}(F_{t})},
\qquad
\pi \;=\; \frac{\max_{t} P_{t}}{\min_{t} P_{t}}.
\end{equation}
Per-frame $F_{t},P_{t}$ trace $\Mdot$, while the centroids $(c_{x,t},c_{y,t})$ trace orbital motion. The second moments $(\sigma_{x,t},\sigma_{y,t})$ trace the bow-shock extent and thus $v_{\infty}$, and $\phi, \pi$ try to pick up periastron-driven temporal variability due to the eccentricity.

We standardize every summary component by its empirical mean $\bar{s}_{k}$ and standard deviation $\hat{\sigma}_{k}$ estimated from a batch of $1000$ test simulations, and use the standardized $\ell_{2}$ distance. The acceptance threshold $\epsilon$ is set adaptively as the $5\%$ quantile of distances measured in a pilot run of $500$ previous simulations. We then sweep the full training pool of $N=40{,}000$ pre-computed $(\bm{\theta}_{i},\mathbf{I}_{i})$ pairs and accept $\bm{\theta}_{i}$ whenever the condition $d\!\left(\mathbf{s}(\mathbf{I}_{i}),\mathbf{s}_{\mathrm{obs}}\right)\le\epsilon$ is met.

\paragraph{Reference-case posteriors}
\begin{figure}[h]
  \centering
  \begin{subfigure}[b]{0.49\linewidth}
    \includegraphics[width=\linewidth]{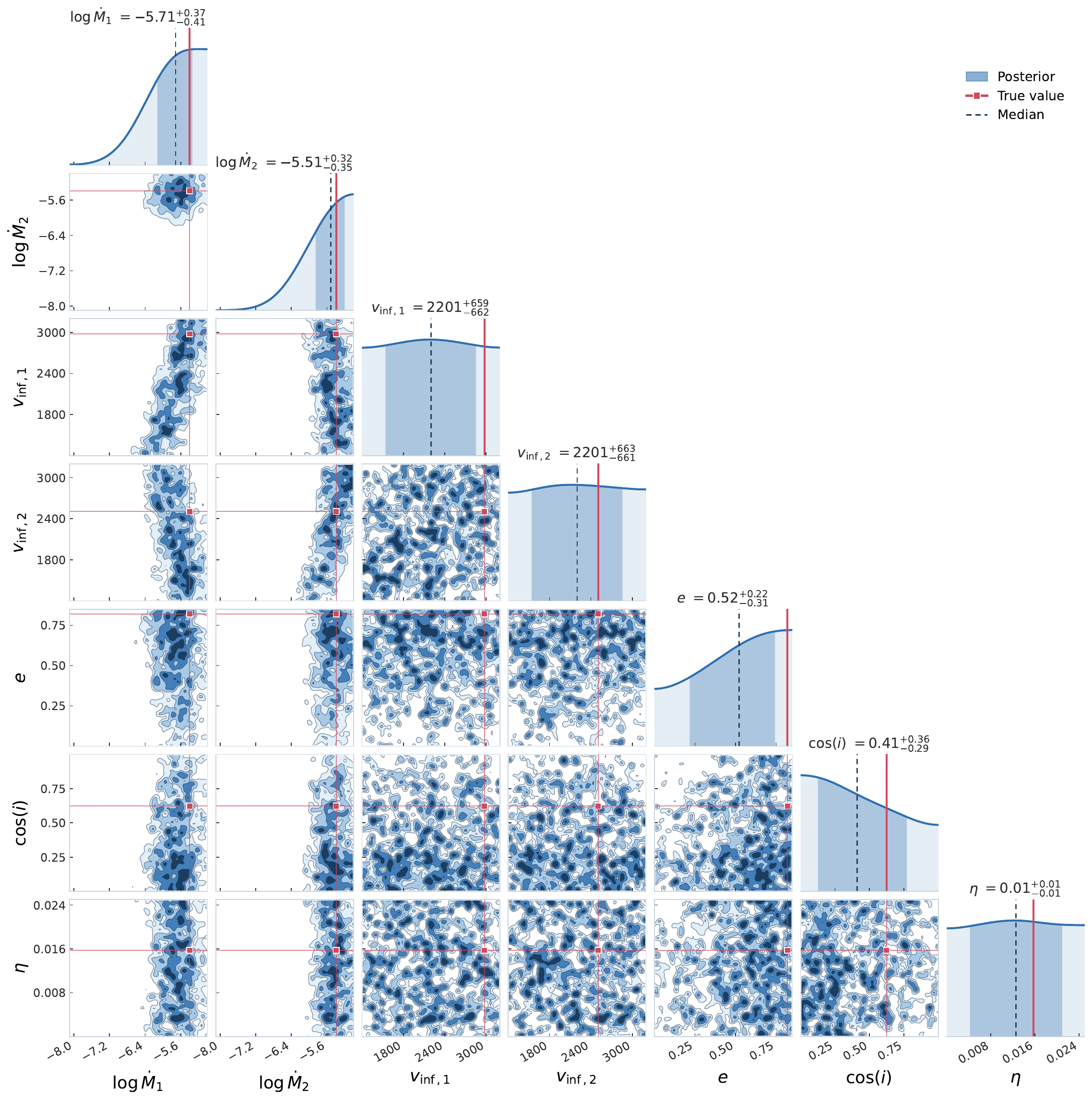}
    \caption{Case~1 (high wind luminosity).}
  \end{subfigure}\hfill
  \begin{subfigure}[b]{0.49\linewidth}
    \includegraphics[width=\linewidth]{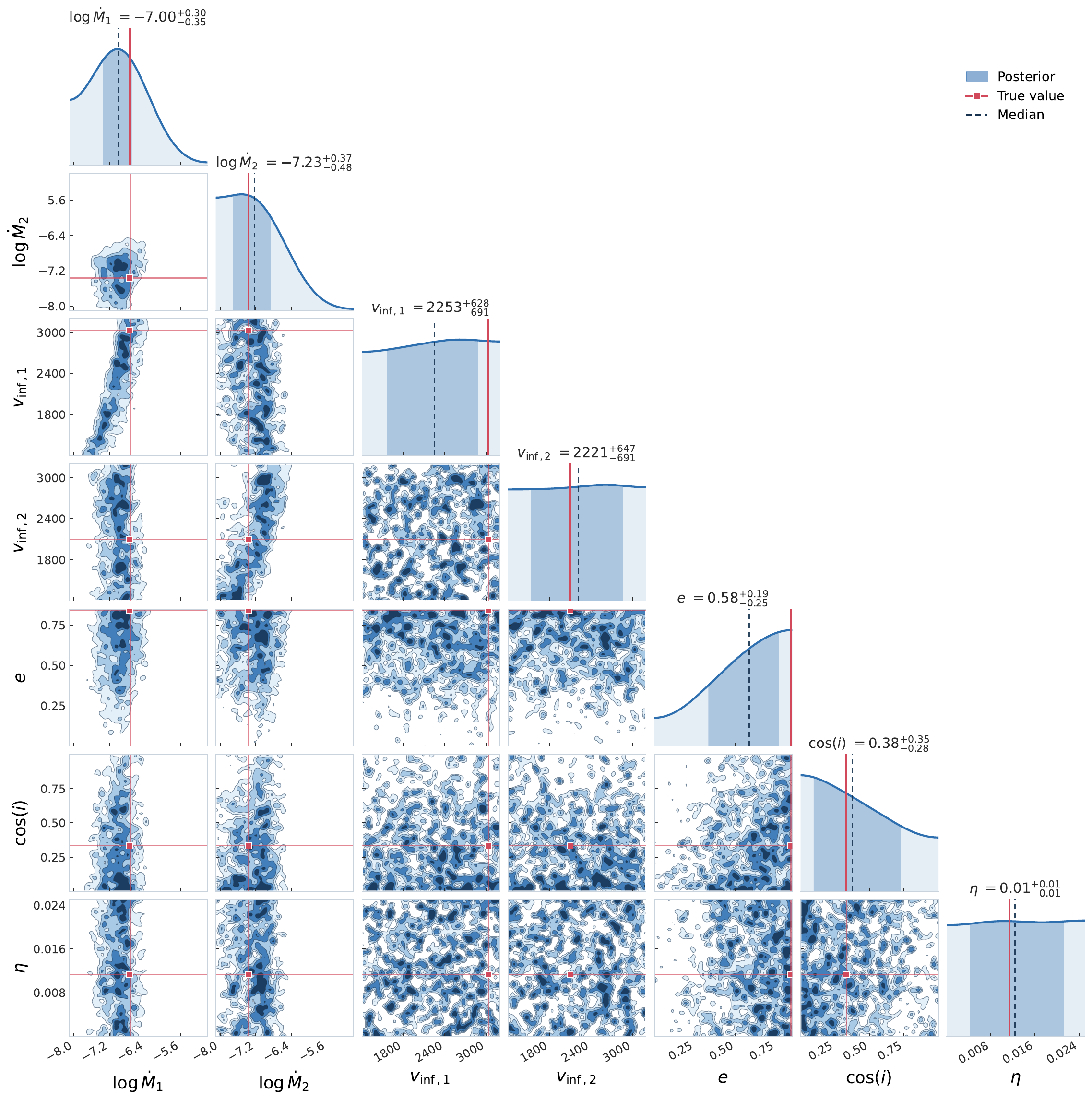}
    \caption{Case~2 (low wind luminosity).}
  \end{subfigure}
  \caption{ABC-rejection posteriors obtained from the $62$-D hand-crafted summary of Eq.~\ref{eq:abc_summary} for the reference cases~1 and~2 (see Tab.~\ref{tab:cases}). Only the mass-loss rates are constrained, with widths noticeably broader ($1\sigma$ CI spans around a full decade for $\Mdot_{1/2}$ in case~1) than those of the neural estimator. All other posterior distributions closely resemble their respective prior distributions.}
  \label{fig:abc_posteriors_app}
\end{figure}

\newpage
\paragraph{Structural limitations of ABC rejection.}
Beyond the poor recovery of kinematic and orbital parameters, the rejection sampler carries three structural drawbacks that motivate the neural approach. First, the rejection step is inherently simulation-wasteful: at a $5\%$ acceptance rate, $95\%$ of the $N=40{,}000$ forward simulations are discarded, so that the effective posterior is built from at most $\sim\!2000$ samples regardless of total compute invested. Second, ABC rejection is not amortized, which means that the full sweep over $N=40{,}000$ simulations must be repeated independently for every new observation, consuming the entire simulation budget per target. Third, global calibration diagnostics such as TARP and SBC (Appendix~\ref{app:calib}) require evaluating posteriors over many held-out observations simultaneously, which is computationally prohibitive for rejection sampling and no population-level calibration statement can therefore be made for the ABC estimates.

\end{document}